\documentclass{elsart5p}
\journal{BioSystems}

\usepackage[authoryear]{natbib}
\usepackage{graphicx}
\usepackage{amssymb}

\begin{document}

\begin{frontmatter}



\title{Transport Characteristics of  Molecular Motors\thanksref{birth}}
\thanks[birth]{Dedicated to Prof. Zbigniew Grzywna on the occasion of his  60 birthday.  
Presented at the ZJG60 Symposium on Bionanomaterials with Structure-Property 
Relationship, 10-11 March, Gliwice, Poland }

\author{Lukasz Machura\corauthref{cor}\thanksref{FNP}},
\corauth[cor]{Corresponding author.}
\ead{Lukasz.Machura@us.edu.pl}
\author{Marcin Kostur}, 
\author{Jerzy \L uczka}
\address{Institute of Physics, University of Silesia, 40--007 Katowice, Poland}

\thanks[FNP]{Supported by the Foundation for Polish Science (FNP)}

\begin{abstract}
Properties of  transport of molecular motors are
investigated. A  simplified model based on the concept of  Brownian
ratchets is applied. We analyze  a stochastic equation of
motion by means  of  numerical methods. The transport
is systematically studied with respect to its energetic efficiency and
quality expressed by an  effective diffusion coefficient. We demonstrate the role  of
 friction and non-equilibrium driving on the transport quantifiers and identify  regions of 
a  parameter space where motors are optimally transported.  
\end{abstract}


\end{frontmatter}

\section{Introduction}
\label{intro}
In the process of evolution Nature has created perfect biomachines.
Researchers (read: visionaries) have dreamed to get ahead of Nature and
construct molecular{}-level machines which would be at the ultimate
limit of miniaturization. Recent development of nanofabrication
technology establishes a foundation for designing, synthesizing,
constructing and testing functional hybrid mechanical and electrical
devices on a nanometre scale. Such nanomachines, operating under
different conditions, could find very many new applications. Nowadays,
there are many problems that have to be overcome: how to power such
systems, actuate and transport them, and couple to another systems.
Nature prompts and provides partial solutions: Within every living cell
there is a service structure for its effective function and operation,
transport and distribution of various products and substances inside,
outside and across cells \citep{How01}. These processes are mediated by tiny
bio-motors that move along filamentous tracks. They do mechanical work
consuming energy taken from the hydrolization of adenosine triphosphate
(ATP) which operates as a power station. The superfamilies of
molecular bio-motors like kinesin, dynein and myosin are responsible for
transport of vesicles and organelles, chromosome segregation, cell
division and muscle contraction, to mention only a few. Although
biological, chemical and mechanical aspects have been studied,
relatively little is known about the mechanisms how the biological motors work.
The fundamental question is whether their movement is purely
deterministic or rather random which only looks like deterministic on a
macro{}-scale.

Biological motors possess many of the characteristics required to power
molecular machines. They can generate force and torque, transport
various cargoes and are able to operate in a processive manner, i. e.
they can move continuously along the specific substrates for distances
of up to hundreds of steps (several microns). An example of such a
motor is conventional kinesin \citep{How01}. The two heads of the kinesin dimer work
in a coordinated manner to move along one of 13 protofilament tracks of
the microtubule. Each protofilament consists of asymmetric $\alpha\beta$--tubulin heterodimers. A heterodimer is about 8 nm long and is
asymmetric because it is composed of two globular subunits: $\alpha$--tubulin
and $\beta$--tubulin which are joined together in a head--to--tail fashion
so that the dimmer has a translational symmetry. Because the $\alpha\beta$--tubulin heterodimers are asymmetrical, the microtubule is
\textit{polar} and its ends are structurally different. One consequence
of this polarity is that polymerization is faster at one end than the
other. The fast--growing end is called the plus--end, whereas the
slow--growing end is called the minus--end. The conventional kinesin
moves towards the plus--end. There are other bio-motors like e.g. ncd dimer \citep{Hen97,Sab98} 
which move in the opposite direction on the same structure. 
The reason, why the motor  moves in a given direction, is not explained to the end. 
 There are several conjectures supported by analysis of very simple  mechanical models.  
One of them says that it can depend on size of the motor, another says that it can depend 
on mass of the motor or more enigmatically it depends on "molecular topology of the motor domain 
relative to the rest of the molecule" \citep{Hen97}.  

Artificial molecular motors are under constructions and experiments on transport properties are performed.
An example is a Brownian motor by using cold atoms in a dissipative optical lattice. In  experiments the optical potential is spatially symmetric and the time symmetry of the system is broken by applying appropriate zero-mean ac forces. As a result, a
current of atoms through the optical lattice can be  generated \citep{Jon04,Sjo06}. 

\section{Model}
\label{model}
From a physical point of view, the kinesin (or other bio-motors) moves in a one dimensional
spatially periodic potential $V(x)=V(x+L)$ of
period $L \approx 8nm$ \citep{Vis99} . The reflection symmetry of this potential
is broken $V(x) \neq V(-x)$, because the
ab{}-heterodimers, that build the microtubule, a highway for the
biomachine, are asymmetrical. This is a crucial fact because it
determines the mechanism of the kinesin movement {--} the ratchet
effect \citep{Ast02,Han05,Luc95}. To explain this mechanism, let us consider as simple
model as possible. We assume that the molecular motor is a particle of
mass $m$ moving in a periodic potential $V(x)$ of period
L and of the barrier height
$\Delta V=V_{max} - V_{min}$. The equation of motion
for the motor is the Newton equation with a complementary random force
which corresponds to thermal fluctuations, i.e. the Langevin equation
in the form 
\begin{equation}\label{LE}
  m\ddot{x}+\gamma \dot{x}=f(x)+g(t)+\sqrt{2D}\Gamma (t)
\end{equation}
Inertial effects, related to the mass,  are described by the first term on the left hand
side. The dissipation is included via the Stokes force with the
friction coefficient $\gamma$ which is proportional to linear size $R$ of
the particle i.e.,
\begin{equation}
  \gamma =6\pi \eta R
\end{equation}
and is additionally determined by the viscosity $\eta$ of the medium
the particle moves in. The potential force
\begin{equation}
  f(x)=-{\frac{dV(x)}{\mathit{dx}}}
\end{equation}
is zero over a period L,
\begin{equation}
  \langle f(x)\rangle _{L}=\frac{1}{L}\int
_{0}^{L}f(x)\mathit{dx}=\frac{1}{L}[V(L)-V(0)]=0.
\end{equation}
The stochastic force $\Gamma(t)$ describes thermal fluctuations which
can be modeled by $\delta$--correlated Gaussian white noise of the
statistics
\begin{equation}
  \langle \Gamma (t)\rangle =0,\ \ \ \ \ \langle \Gamma (t)\Gamma
(s)\rangle =\delta (t-s).
\end{equation}
According to the dissipation--fluctuation theorem, the thermal noise
intensity $D$ is related to the friction coefficient and temperature $T$
of the system, i.e.,
\begin{equation}
  D=\gamma k_{B}T,
\end{equation}
where $k_B$ stands for the Boltzmann constant.

The external time--dependent force $g(t)$ can be of any genre,
both deterministic or stochastic \citep{Luc99,Kul98}. This force makes the  system to be in 
a non-equilibrium state 
and is a source of energy for movement of motors. For biological motors it comes from
chemical reactions. Here, as an example and for the didactic purposes,
we choose a time periodic force, namely
\begin{equation}
  g(t)=A\cos (\Omega t),\label{g}
\end{equation}
where $A$ is the amplitude and $\Omega$ stands for the frequency of the
external stimulus. Equivalently we can define the period $T = 2 \pi/ \Omega$
of the external force. This kind of force can be realized for artificial motors while for biological motors, random  non-equilibrium force is more adequate.

%%%%%%%%%%%%%%%%%%%%%%%%%%%%%%%%%%%%%%%%%%
\subsection{Time scales}

In physics only relations between scales of length, time and energy are
relevant, not their absolute values. Therefore we shell now translate
the above equation of motion into its dimensionless form. First of all
we determine characteristic quantities -- time and length. The characteristic length is  
 the period L of the potential $V(x)$ and accordingly the coordinate of
the molecular motor can be scaled as
\begin{equation}
  y=\frac{x}{L}.
\end{equation}
Time can be scaled in several ways. One of the possibilities is the
relaxation time $\tau_L$ of velocity or the correlation time
of the velocity of the massive Brownian particle moving freely and
driven only by the thermal noise. It can be extracted from a special
case of Eq. (\ref{LE}) when the right hand side is zero, i.e.,
\begin{equation}
  m\dot{v}+\gamma v=0. \label{NE0}
\end{equation}
From the above equation it follows
\begin{equation}
  v(t)=v(0)\ \exp (-t/\tau _{L}),\label{vel}
\end{equation}
where the characteristic time $\tau_L = m/\gamma$ is
sometimes called the Langevin time. Another characteristic time comes
from the overdamped motion of the particle in the potential $V(x)$, when
Eq. (\ref{LE}) reduces to the form
\begin{equation}
  \gamma \frac{\mathit{dx}}{\mathit{dt}}=-\frac{\mathit{dV}(x)}{\mathit{dx}}. \label{NE1}
\end{equation}
Then, by inserting into above equation the characteristic quantities, we
get the definition of time $\tau_0$, 
\begin{equation}
  \gamma \frac{L}{\tau _{0}}=\frac{\Delta V}{L},\ \ \ \rightarrow
  \ \ \ \tau _{0}=\frac{\gamma L^{2}}{\Delta V}. \label{t0}
\end{equation}
During the time interval $\tau_0$ overdamped particle proceeds
the distance $L$ under the influence of the constant force $\Delta V/L$. Third
characteristic time follows from the friction less equation of motion,
i.e., when Eq. (\ref{LE}) takes the form
\begin{equation}
  m\frac{d^{2}x}{\mathit{dt}^{2}}=-\frac{\mathit{dV}(x)}{\mathit{dx}}. \label{NE2}
\end{equation}
From the above equation the characteristic time $\tau_m$ is
given by the relations
\begin{equation}
  m\frac{L}{\tau _{m}^{2}}=\frac{\Delta V}{L},\ \ \ \rightarrow
  \ \ \ \tau _{m}^{2}=\frac{mL^{2}}{\Delta V}. \label{tm}
\end{equation}
One can distinguish also other characteristic times like time period of
the external driving $T$ or the well known Einstein diffusion time
\begin{equation}
  \tau _{E}=\frac{L^{2}}{2D_{E}},\ \ \ D_{E}=\frac{k_{B}T}{\gamma }. \label{td}
\end{equation}
Now we can rescale the equation of motion for the massive Brownian
particle in several ways. Doing this we shell take as the relevant time
scales these times which differs for different systems, like
$\tau_m$ or $\tau_0$. Let us note that the Langevin
time and the Einstein diffusion time do not depend on the system
itself, i.e., on the potential and the external driving forces.

%%%%%%%%%%%%%%%%%%%%%%%%%%%%%%%%%%%%%
\subsection{Rescaled equations of motion}

Let us first propose $t_{0}$ as the relevant time scale,  i.e. the rescaled dimensionless time is 
$s=t/\tau_0$. The dimensionless form of the
Langevin equation then reads  
\begin{equation}
  \varepsilon \ddot{y}(s)+\dot{y}(s)=F(y)+G(s)+\sqrt{2D_{0}}\xi (s), \label{LEs1}
\end{equation}
where dot denotes derivative with respect to the rescaled time
$s$, $F(y) = - dW(y)/dy =-W'(y)$ denotes the rescaled potential force
and $G(s) = (L/\Delta V)g(t) = a\cos(\omega s)$
stands for the rescaled external driving force with the
rescaled amplitude $a=(L/\Delta V)A$ and frequency 
$\omega = \Omega \tau_0$. The
rescaled spatially periodic potential 
$W(y) = V(x)/ \Delta V = V(Ly)/ \Delta V = W(y+1)$ 
has the unit period and the unit
barrier height. The dimensionless mass
\begin{equation}
  \varepsilon =\frac{m}{\gamma \tau _{0}}=\frac{\tau _{L}}{\tau _{0}} \label{mass}
\end{equation}
is a ratio of the two characteristic times. The rescaled thermal noise
$\xi(s) = (L/\Delta V)\Gamma(t) = (L/\Delta V)\Gamma(\tau_0 s)$ has exactly the same statistical
properties as $\Gamma(t)$. The dimensionless noise intensity
\begin{equation} \label{D0}
D_0 = \frac{k_B T}{\Delta V}
\end{equation} 
 is a ratio of thermal
energy to activation energy the particle needs to traverse the
non--rescaled potential barrier.

On the other hand, if we choose the dimensionless time  $u=t/\tau_m$ and $\tau_m$ as a time scale 
then we end up with another version of the rescaled Langevin
equation, namely
\begin{equation}
  \ddot{y}(u)+\hat{\gamma }\dot{y}(u)=F(y)+G(u)+\sqrt{2\hat{\gamma}D_{0}}\xi (u). \label{LEs}
\end{equation}
The dimensionless friction coefficient is a ratio of two characteristic
times, different then previous two, namely,
\begin{equation}
  \hat{\gamma }=\gamma \frac{\tau _{m}}{m}=\frac{\tau _{m}}{\tau _{L}}. \label{friction}
\end{equation}
The rescaled driving force $G(u) = a\cos(\omega u)$
with the rescaled amplitude $a=(L/\Delta V)A$ and frequency
$\omega = \Omega \tau_m$. 

Two different scaling are useful in two limiting regimes: Eq. (\ref{LEs1}) in the overdamped case 
(when  $\varepsilon <<1$ is a small parameter) while 
Eq. (\ref{LEs}) in the underdamped case (when $\hat{\gamma} <<1$ is a small parameter).

\subsection{Numerical values for kinesin}

Let us evaluate  characteristic times for one of the best known biological motor, namely,  kinesin which moves along a microtubule.  As already mentioned in the Introduction, microtubules are spatially
periodic structures of period $L \approx 8nm$. The mass of the kinesin head
domain is of order
$m=100kDa=1.66 \cdot 10^{-22}kg$
and its radius is  $R=3nm$. The friction coefficient
$\gamma = 6 \cdot 10^{-11} kg/s$ is
calculated from the Stokes formula with the use of the viscosity of
water ($\eta = 10^{-3} kg/ms$). In a typical Brownian
domain the activation energy is 5 time higher than the thermal energy,  
$\Delta V = 5 k_B T$ and the temperature inside cell is  about $310K$
($37^{o} C$). W can now estimate the typical characteristic
times for the kinesin moving inside human cell; thus
\begin{eqnarray}
  &\tau_{L}&=2.77\cdot 10^{-12}s,\ \ \ \tau _{0}=1.8\cdot
10^{-7}s, \nonumber \\
&\tau_{m}&=7\cdot 10^{-10}s,\ \ \ \tau _{E}=4.57\cdot
10^{-7} s. \nonumber
\end{eqnarray}

The dimensionless mass (\ref{mass}) in the first scaling and the dimensionless
friction (\ref{friction}) in the second scaling have the following values
\begin{eqnarray} \label{licz}
  \varepsilon =1.54\cdot 10^{-5}\ll 1,\ \ \ \gamma =2.5\cdot 10^{2}. \nonumber
\end{eqnarray}

One can note that the value of the parameter  $\varepsilon$ is very small.  Therefore 
 Eq. (\ref{LEs1}) seems to be more appropriate than Eq. (\ref{LEs}) because
(\ref{LEs1}) contains the  small parameter $\varepsilon$. This allows, with a very good approximation, to put formally $\varepsilon = 0$ in the dimensionless equation of
motion (\ref{LEs1}) yielding 
\begin{equation} \label{over2}
\dot{y}(s)=F(y)+G(s)+\sqrt{2D_{0}}\xi (s). 
\end{equation}
Analysis of this equation is much easier than Eq. (\ref{LEs1}).  From (\ref{licz}) it follows that 
the case of overdamped  dynamics  takes place for biological motors. 
On the other hand, Eq. (\ref{LEs}) contains the parameter $\hat{\gamma}$ 
and one cannot find justifiable arguments to neglect any term in it.  So, this equation is much harder to analyze in a complete manner. 
However, if one wants to investigate some particular effects like the influence of the
inertia on the transport,  Eq. (\ref{LEs})  is more practicable. 
For other motors, especially non-biological,  inertial effect can be crucial. 
An example is an atomic Brownian motor moving in optical lattices \citep{Brown08} and Eq. (1) in 
\citep{Hag08}. 
It is worth to stress that if  the non-equilibrium driving is of the form (\ref{g}), the dynamics can be chaotic in some 
regimes leading to anomalous transport behavior like negative mobility or negative conductivity \citep{Mac07,Kos08}.  
Below, we present characteristics of transport in full regime, from underdamped  to overdamped one.  
 From now on we will use the later scaling
(\ref{LEs}) while defining performance characteristics of the motion of the
Brownian motor and shell omit all the hats in Eq. (\ref{LEs}), for the sake of
simplicity.

%%%%%%%%%%%%%%%%%%%%%%%%%%%%%%%%%%%%
\section{Performance characteristics of molecular motor} 

When one study the motion of molecular motors, the most important  transport
measure is an average stationary  velocity
$ \langle\: v \:\rangle$ of the 
motor \citep{Mac04}. 
Averaging should be performed over all realization of thermal noise,
initial conditions  and over a period of the external driving. Average velocity  
describes how much time a typical particle needs to overcome a given
distance in the asymptotic (long--time)  state. This  average velocity,
however, is not the only transport attribute. Other
characteristics  are also important. We will analyze  two following transport 
aspects:  quality  and energetic efficiency.    The  quality of transport  can be characterized 
by the effective
diffusion coefficient $D_{eff}$, i.e., by the fluctuations
in the position space  \citep{Lin01}
\begin{equation}
  D_{\mathit{eff}}= \frac{\langle x^{2}\rangle -\langle x\rangle^{2}}{2t}, \label{Deff}
\end{equation}
where the brackets $\langle \dots \rangle$ denote averaging over
all realizations of thermal noise and initial conditions. 

The third  quantifier is the efficiency of noise rectification \citep{Mac04,Suz03}. The
motor moves in  viscous media.  Therefore the minimal energy input required to move a particle
in presence of friction $\gamma$ over a given distance depends on the
velocity. In this case, the rectification efficiency is given by the
formula \citep{Mac05,Suz03,Lin05,Kos06}
\begin{equation}
  \eta =\frac{\gamma  \langle v\rangle^{2}}{P_{in}}. \label{eta}
\end{equation}
The average input power $P_{in}$ corresponding to system described by Eq. (\ref{LEs})
 is given by the formula 
\begin{equation}
  P_{in}=\gamma [ \langle v^2 \rangle -D_{0}], 
\end{equation}
where $D_0$ is defined in Eq. (\ref{D0}).  
This expression follows from the energy balance of the underlying
equation of motion (\ref{LEs}) \citep{Mac04}. 

%%%%%%%%%%%%%%%%%%%%%%%%%%%%%%%%%%%%%
\section{Numerical experiment and discussion}

To examplify the above ideas we analyze  Eq. (\ref{LEs}) in the long-time asymptotic,
time--periodic regime after effects of the initial conditions and
transient processes have died out. Then, the statistical quantifiers of
interest can be determined in terms of the statistical average over the
different realizations of the process (\ref{LEs}) and over the driving period
$T$. Since there are no analytical methods to handle (\ref{LEs}), we made use of
numerical approach and have carried out extensive and precise numerical
simulations, applying Stochastic Runge--Kutta algorithm of order 2. 
For the initial conditions we took points from uniformly distributed circle 
in the phase space $(x,v)$  with the radius $r = \sqrt{x^2 + v^2} = 1$ and the origin in 
the point  ($x_{min}$, 0),
where $x_{min}$ denotes  minimum of the potential $W(y)$.
 For the illustration of the above idea we  fixed  the
following set of parameters:  $\omega = 4.9$, $D_0 = 0.001, 0.05$
and the potential profile $W(y) = \Delta W [\sin(2 \pi y) +
0.25 \sin(4 \pi y)]$, where $\Delta W = 0.454$ reduces the maximal
barrier height to unity. The forces corresponding to this potential
ranges from the minimal value -2.14 to  the maximal value 4.28. If the  amplitude $a$ of the
driving is higher than 4.28,  the  motor 
is able to overcome the potential barrier in any direction just with
the use of the external driving force and without of thermal noise.
\begin{figure}[ht]
  \begin{center}
    \includegraphics[angle=0,width=0.9\linewidth,clip=]{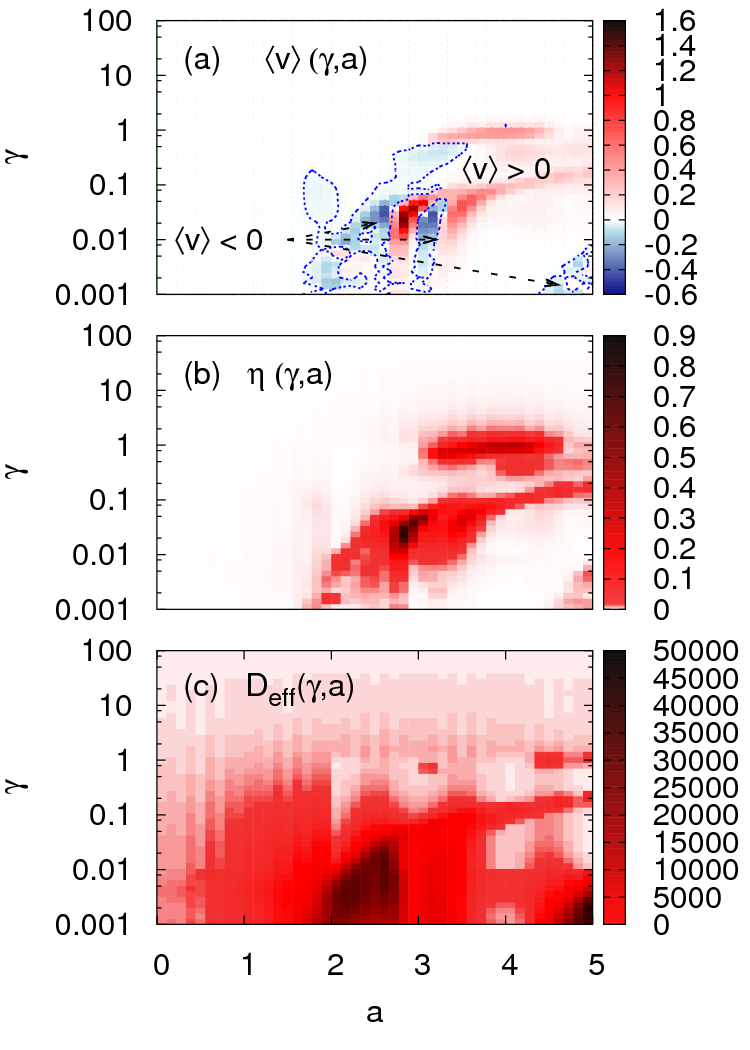}
  \end{center}
  \caption{(color online) Performance characteristics of the molecular motor are 
presented versus the external force  amplitude  $a$ and friction coefficient $\gamma$.
  In  panel (a) we show the average velocity $|\langle v \rangle|
  <10^{-2}$ (white), negative (blue, gray surrounded by contour), positive (red, gray).
  In  panel (b) we present the  efficiency. In  panel (c) we depict the 
  effective diffusion. 
  Other parameters are: $D_0=0.001$, $\omega=4.9$.
  }
  \label{fig1}
\end{figure}
We analyze the system (\ref{LEs}) in the  2-dimensional 
parameter space $\{\gamma, a\}$.  We change the rescaled friction coefficient $\gamma$  from
$\gamma=0.001$ (underdamped dynamics)  to 
$\gamma=100$ (overdamped dynamics).  The external time periodic force strength
$a$ varies between 0 and 5, where the later is just above the maximal value of the 
potential force.

In Fig. \ref{fig1}  we show three  characteristics  in the 
 low temperature  regime,  $D_0 = 0.001$, cf. Eq. (\ref{D0}).
In  panel (a),  we depict  the stationary average  velocity of the  motor. 
White area  corresponds  to the case when the absolute value of the  velocity is small,  
$|\langle v \rangle| <10^{-2}$.
Blue (or gray) areas surrounded by contours stands for the negative mean velocity and
red (gray) color indicates regions of  positive velocity. It is  seen that for a chosen
set of parameters the Brownian motor has  a noticeable  velocity for the  force  amplitude 
$a > 2$ and for  not strong damping $\gamma < 2$. Strictly speaking,  the velocity  is
not exactly zero  assuming very small values of order $10^{-2} \div 10^{-7}$ 
(white area on Fig. \ref{fig1} (a)).
By inspecting the several colored (gray) areas, one can observe that the velocity as a function of $\gamma$ or the amplitude $a$ exhibits the  multiple  velocity  reversals
\citep{Kos01}.  Because the friction coefficient $\gamma$ depends on the linear size of the motor, it means that motors of different sizes can move in  opposite directions. 
In panel (b) we show the rectification efficiency (\ref{eta}).  There, white area corresponds to 
efficiency smaller that  $\eta< 10^{-3}$. The correlation of dark regions  in panels (a) and (b) 
is evident. 
There are two main islands of efficient energy conversion -- one
 for  $\{\gamma, a\} \in \{0.4\div2, 3\div5\}$ and second for 
$\{\gamma,a\} \in \{0.001 \div 0.2, 2\div5\}$.
In the region of parameters $\{\gamma \approx 0.002,  a \approx 2.9\}$, the  motor efficiency 
is very high and is almost  90\% (remember that the system is far from an equilibrium state).  
In  panel (c) we show the effective diffusion coefficient of the  motor.
Large value of this quantifier means low quality of motion - Brownian particles
move in a very irregular manner. 
This  region of high efficiency and low quality of motion (high diffusion), 
makes our stochastic model similar 
to  deterministic \citep{Mat00} or Hamiltonian \citep{Sch01} systems.

Upon the inspection of Fig. \ref{fig1} one can notice that for the parameter island
with large {\it negative} average velocity and high efficiency (large area around
$\gamma=0.02$ and $a=2.8$),  the motor moves quite 
irregularly  because  $D_{eff}$ is large. For the island with large {\it positive} velocity and 
 medium efficiency (smaller area around $\gamma=1$ and $a=4$),
the effective diffusion is small reflecting  good quality of transport and
regular motion.

\begin{figure}[htbp]
  \begin{center}
    \includegraphics[angle=0,width=0.9\linewidth,clip=]{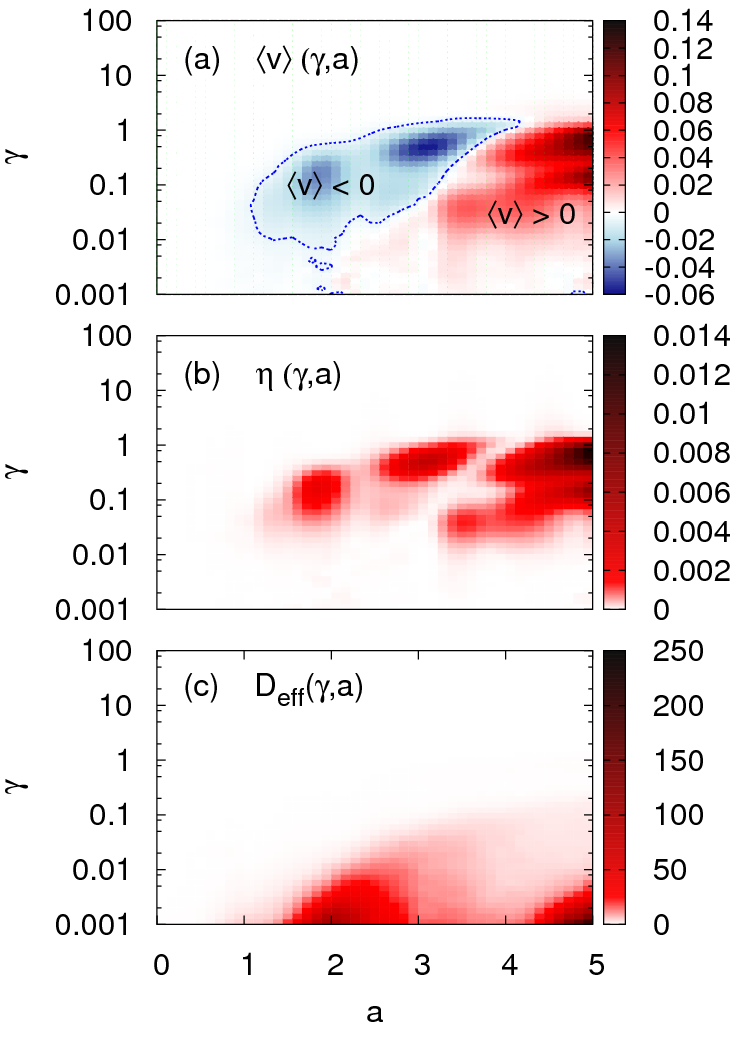}
  \end{center}
  \caption{(color online) The same as in Fig. 1 but for higher temperature, $D_0=0.05$. 
 Panel (a): the average velocity $|\langle v \rangle|
  <10^{-3}$ (white), negative (blue, gray surrounded by contour), positive (red, gray).
  Panel (b):   the  efficiency.  Panel (c):  the effective diffusion. 
  }
  \label{fig2}
\end{figure}

Let us  consider the case of higher temperature, $D_0=0.05$.  Transitions of a particle 
over  potential barriers  are easier  and details of the shape of the potential is not so important now. 
As one can see from  panel (a) of Fig. \ref{fig2},    absolute values of the average velocity
are one order  smaller  than in the previous temperature regime. Higher temperature makes
the motor motion   slower in both negative -- blue (gray) contoured areas, or positive 
-- red (gray) colored regions. Again as in the previous case,  in the the  2-dimensional 
parameter space $\{\gamma, a\}$,  we can identify the velocity reversal phenomenon upon 
the change  of the control parameters $a$ or $\gamma$. 
In the negative direction  the maximal velocity  $\langle v\rangle =-0.06$  and in the positive direction   $\langle v\rangle=0.14$.
In the optimal regime, the efficiency of the motor transport  
is of order  1\%. The only characteristic which seems to be better  
than in the previous scenario is the effective diffusion presented in  panel (c). 
The motor  seems to move in a much more regular manner than for lower temperature 
yielding maximum value of $D_{eff}=250$ for underdamped system. It is a  region 
where the averaged velocity  is almost zero and the 
efficiency is very low.

Comparison of Figs  1 and 2 leads to the conclusion that the influence of higher temperature  
is rather destructive. The only better quantifier is the effective diffusion coefficient which is smaller. 
It means that  the islands in the parameter space $\{\gamma, a\}$  for higher  temperature with relatively high (negative or positive) velocity 
accompanied by the highest possible efficiency represent regimes where the motor  moves
rather in a regular way.

\section*{Conclusions}
The most demanded properties of any transporting machinery are:
efficiency and quality. The latter in our case is characterized by the
effective diffusion coefficient. 
Let us note that even in any equilibrium system the particle can be
transported over long distances due to thermal diffusion. It is,
however, very unreliable when the distance becomes large because the
diffusion cannot distinguish direction and most of traveling particles
would not arrive at prescribed destination in a reasonable time. On the
other hand, from Figs \ref{fig1} and \ref{fig2} we can see that the 
large diffusion is sometimes  in regimes where the energetic efficiency is
high. In real situations, Nature chooses between above scenarios.
Under some circumstances, if e.g. the distances are small, the
diffusion can be exploited  to transport the cargo and no energy is wasted. When the
distance is larger, then the non--equilibrium transport is applied and
reliability is achieved sacrificing the energy input.

\section*{Acknowledgments}
The  work supported in part by the 
Polish Ministry of Science and Higher Education (Grant No. 202
131 32/3786).
One of the authors (LM) gratefully acknowledges financial support by the 
Foundation for Polish Science (FNP).
%The authors gratefully acknowledge financial support by
%the ESF Program ``Stochastic Dynamics: Fundamentals and Applications''.

\end{document}